# Assembly Spaces: Formal Definitions and Fast Methods for Approximating Assembly Indices

Gage Siebert[1], Redwan Chowdhury[2,3], Louie Slocombe[2], Sara Walker[1,2,4,*]

[1]*School of Earth and Space Exploration, Arizona State University*
[2]*Beyond Center for Fundamental Concepts in Science, Arizona State University*
[3]*Siebel School of Computing and Data Science, University of Illinois Urbana-Champaign*
[4]*Santa Fe Institute, Santa Fe, New Mexico, United States*
**author for correspondence: sara.i.walker@asu.edu*

## Abstract

Assembly theory provides a framework for detecting life across diverse substrates. Its two central observables are assembly index and copy number. The assembly index is the minimum number of joining operations needed to construct an object from its elementary parts; for molecules, it can be measured using mass spectrometry, infrared spectroscopy, and NMR. Copy number is the abundance of a distinguishable object within a sample. High assembly index combined with high copy number constitutes a signature that cannot arise abiotically, validated experimentally for molecular biosignatures. A foundational underlying concept is the assembly space, which encodes the causal possibilities determinable from observed objects, with the assembly index the shortest path to them given a substrate's physical constraints. Here, we provide a generalized formalism for assembly spaces and tools for approximating assembly indices. We review applications across molecules, minerals, and atmospheres, then introduce general, substrate-independent definitions of assembly spaces and indices. We develop a unified path hierarchy to clarify relationships among their representations in the molecular assembly literature. Finally, we show how formal grammar algorithms can efficiently bound assembly index calculations, increasing the accessibility of this emerging area across chemistry, biology, and complexity science.

# Introduction

The central challenge in detecting signatures of life is that there is no definition of life general enough to apply beyond known examples[1]. Most life detection frameworks rely on identification of molecules known to be produced by Earth's biology, such as oxygen, methane, or specific organics, and are ambiguous due to abiotic sources. Furthermore, life elsewhere may be chemically very different than what evolved on Earth. Even across Earth's biosphere it is not readily apparent how to identify the imprints of life and evolutionary processes in substrates beyond the biochemically obvious, for example in artificial life[2]. What has been missing is a framework that allows unambiguous identification of the signatures of life based of universal physical principles, without presupposing the specific chemical instantiation of any known living form.

Assembly theory (AT) was developed to address this challenge. Its starting point is a simple but potentially powerful observation: complex molecules found in high abundance are candidate universal biosignatures. As the structural complexity of a molecule increases, the number of possible molecules at that complexity level grows at least exponentially[3,4]; one might then expect the chance that any specific complex molecule arises and persists in detectable abundance by undirected chemistry alone to decreases rapidly with complexity[5]. A sufficiently complex molecule detected in high abundance thus provides evidence of a process capable of repeatedly and preferentially producing it, e.g., a process of selection, in the broadest sense, characteristic of living systems. To develop a general theory for life, the challenge is then to turn this observation into a rigorous, experimentally verifiable framework: one must define what "complexity" means in a way, that is, in principle, substrate-dependent and, in practice, metrologically accessible.

AT approaches this by grounding complexity in physical causation as opposed to using description-length based complexity in an arbitrary formalism. Rather than asking how compactly an object can be described using a 'simple' universal reference machine, an approach widely adopted in algorithmic complexity, AT instead asks: what is the minimum number of physical operations that must have been performed to produce this object from simpler parts? This leads to the definition of a physical quantity, the *assembly index*, which provides a measurement of causal depth in the size of the shortest recursive path consistent with an object's physical constitution[6]. Crucially, by reframing the question as one of physics and minimum causation, one can develop experimental approaches to answer it. For molecules, the assembly index can be determined from mass spectrometry (MS), nuclear magnetic resonance (NMR) spectroscopy, and infrared (IR) spectroscopy[7], making it the first intrinsic, experimentally tractable measure of molecular complexity[8]. This metrological foundation distinguishes the assembly index from the many topological or graph-theoretic[9–11], and information-theoretic[12,13] complexity measures that have been proposed across chemistry and biology: those measures encode mathematically well-defined properties of molecular structure, but none had an experimental referent, nor an ontology that makes explicit why the property should be considered intrinsic to the molecule rather than an artifact of the chosen complexity measure.

The second central observable in AT is *copy number*: the abundance of a given distinguishable object in a sample. The combination of high assembly index and high copy number carries strong evidential weight for selection. High copy number rules out singular or rare events: something must be capable of producing the object repeatedly. Together, assembly index and copy number define an assembly threshold: a boundary in assembly space such that objects above it cannot be observed in high copy numbers abiotically[14]. For molecular systems, this threshold has been empirically determined: molecules with a molecular assembly index (MA) > 15, in copy numbers detectable by tandem mass spectrometry (> 10,000 copies), were exclusively biogenic in a dataset spanning biological samples, meteorites, simple chemical mixtures, and blinded unknowns [5]. This is the first theoretically hypothesized molecular biosignature that has been validated metrologically.

Further development of AT, and in particular testing whether the framework can apply to substrates beyond molecules, requires continued development of its formalisms, specifically the concept of an *assembly space*. The assembly space is a substrate-specific structure that defines the possible objects, the elementary units from which they can be built, and the joining operations that determine how those units and their products can be recursively combined to form more complex structures. The assembly space is the formal structure that makes AT's claims precise. It encodes the possible causal hierarchy of a substrate: which objects can, in principle, exist, which assembly pathways lead to them, and the shortest causal path to each object. The assembly index of an object is defined via the assembly space of its substrate: it is the length of the shortest assembly path terminating on that object. Different substrates encode different causal mechanisms and therefore demand different assembly spaces. The construction of an appropriate assembly space is a prerequisite for applying AT to any new domain.

Despite the central role of assembly paths in AT, the concept has been developed and represented in different ways across the literature[5–7,15,16], and these representations have not yet been formally unified or systematically related to one another. Multiple notations for assembly paths appear in published work, reflecting different choices about what information to record and how to represent the causal structure of construction. A correspondence between assembly paths and formal grammars has also been noted[17,18], but has not been fully developed or generalized, limiting its utility. Grammars could be implemented to provide efficiently computational bounds for assembly index calculations, an application that is especially important because the computational problem of finding a shortest assembly path is known to be NP-hard[19]. Adapting algorithms for efficient computation to bound the assembly index can increase the accessibility of this emerging area to a broader group of researchers. The present paper addresses these gaps.

Our contributions are as follows. We begin, in Section 2, by reviewing how the assembly index was metrologically derived for molecules, tracing the path from experimental mass spectrometry to the formal concept of the assembly space and establishing molecular assembly index as a physical observable that is measurable by multiple spectroscopic techniques. In Section 3, we introduce a general, substrate-independent formal definition of assembly spaces and assembly indices, then instantiate this definition for two specific examples, molecules as a physical substrate and strings as a more abstract example. In Section 4, we examine the different representations of assembly paths appearing in the literature and articulate the formal relationships among them, placing them into a lattice, which provides a unified hierarchy for comparing and interpreting different assembly space representations. In Section 5, we provide a brief review of relevant concepts in formal grammars. In Section 6, we establish a precise correspondence between the substrate-independent definition of assembly spaces and classes of formal grammars and provide examples of how substrate-embedded computational frameworks can be developed in the form of context-free grammars for string assembly spaces and hyperedge replacement grammars for molecular assembly spaces. In Section 7, we use this correspondence to implement efficient algorithms for bounding the assembly index across a variety of use cases. These bounds reliably track assembly index while remaining far less expensive than more exact algorithms, allowing effective approximations for large enough assembly indices where exact calculations become intractable. We explain some caveats in using these approximations, particularly for lower bounds which could introduce misclassification of the assembly threshold if not implemented with care.

Together, these contributions aim to provide the community with a self-contained reference for the physical and formal foundations in the definition of assembly spaces, clarify the relationship of assembly index to other formalisms, and in the process to substantially widen the applicability of AT across molecular[16,20,21], evolutionary[22,23], philosophical[24,25], physical[14,26] and technological domains[27,28] in which it has begun to find application.

# Assembly Theory and its Metrology

A recurring pattern in the history of science is in how advances in metrology do not merely confirm existing theories but actively allow generating new ones[29]: metrology can transform what were previously philosophical distinctions into empirical questions. Traditionally, the search for alien life has been hard because we do not know what signatures are unique to life, or if it is possible to measure "life" at all[1]. Thus, researchers in astrobiology and adjacent fields concerned with the definition of life and its application to biosignatures have focused primarily on philosophical debates[30–33] without the support of empirical tests.

AT began with an effort to address the missing gap in metrology for life detection. The idea was that mapping the complexity of molecular space would make it possible to place molecules on a standardized metrological scale: the range of this scale would be from molecules able to form abiotically to those so complex they require a vast amount of encoded information or memory to produce their structures, so much so that they can only be produced in abundance after a process of selection and evolution (e.g., "life"). To build a testable framework, the proposed scale needed to be experimentally verifiable, and this was accomplished by building a model to corroborate theory with spectroscopic data[5].

The resulting theory provides a causation-based account of physical complexity: objects are characterized not by their current state, but by the minimum causation required to produce them, measured as an intrinsic property of objects. For background on the philosophical foundations of AT, see Walker and Cronin[14], Walker[34] and Ardoline[24], on the theory, see Sharma *et al*.[6], for the metrology, see Jirasek *et al*.[7], and for application to life detection and origins of life, see Marhsall *et al*.[5]. Here we review some of the current literature before moving to generalized definitions.

Applying the framework of assembly theory to a substrate, such as for molecules[5], minerals[35], or atmospheres[36], requires formalization of several relevant theoretical concepts. Central is that of an ***assembly space***: a formalism that represents the possible objects, the basic units from which they can be built, and the rules that allow their construction[15]. The definition of the assembly space is determined by joining operations: substrate-specific constraints that determine how two structures can combine to make new structure. Joining operations define possible causal steps, and the assembly space characterizes a causal hierarchy, providing a formal language to underly AT's hypothesis complex objects do not form abiotically, see Walker and Cronin[14]. Sequences of these joining operations define ***assembly paths***. An object's ***assembly index*** measures how many joining operations are needed to construct it from the basic units. For molecules, this can be determined metrologically from mass spectrometry, infrared spectroscopy, and nuclear magnetic resonance[7].

Another critical quantity in assembly theory is ***copy number***, defined as the number of identical instances of an object present in a sample. Copy number is important for two reasons. First, a high abundance of an object is typically required for it to be detectable. Second, a large copy number drives the probability that a given object was formed without selection, by fluctuation, towards zero. A high assembly index object with a high copy number constitutes strong evidence of an evolutionary lineage preceding it, and in the theory this is quantified in terms of Assembly, $A$, where $A = \sum_{i,n_i>1} e^{a_i} \; n_i/N_T$ , where $i$ indexes the distinguishable types of observed objects, and $N_T$ is the total number of objects of any type observed[6]. We refer readers to Sharma *et al.*[6] and Walker and Cronin[14] for an in depth discussion of the theoretical framework of AT. Herein, we do not focus on the full scope of the theory but instead focus on concise that formalize the concepts of assembly space and the assembly index and allow their calculation. To establish the foundations of these concepts formally, we first review the physical motivations for their definitions, as derived from measurement.

## Spectroscopic Measurements of Molecular Assembly Index

The assembly index was first defined through thought experiments regarding how mass spectroscopy interrogates molecular structure. The key observation was that a molecule with a high assembly index (e.g. one requiring many recursive construction steps) would produce a fragmentation spectrum with many distinct peaks in tandem mass spectrometry ($MS^2$), because the structural complexity that requires many construction steps will correspondingly generate many distinct fragments upon dissociation[5]. Conversely, a low-assembly-index molecule, which contains fewer distinct structural motifs, generates proportionally fewer fragment ions.

Mass spectrometry fragmentation allows the association of the number of $MS^2$ peaks and the assembly index[5] using data-dependent acquisition to fragment the most abundant ions in each sample. For the experimental data presented in Marshall et al.[5], no abiotic samples included molecules with $a > 15$, whereas the biological and technological samples included molecules above this threshold value. Measurements of the assembly index have also been developed using infrared (IR) and nuclear magnetic resonance (NMR) spectroscopy. The physical basis for an IR measurement relies on structure in a molecule's fingerprint region (400–1500 $cm^{-1}$) where absorption bands arise from collective vibrational modes and bond-bending motions: these reflect the global structure of the molecular graph rather than isolated functional groups[7]. Since the number of distinct (non-repeated) structural subunits increases with assembly index, by definition, a high-assembly-index molecule contains more distinct, non-reducible motifs; the number of resolvable peaks in the fingerprint region can be directed related to molecular assembly index. For NMR, the physical basis of the measurement is the sensitivity of carbon chemical shifts to local bonding environment: in a molecule with many distinct structural motifs (e.g., high assembly index), the number of chemically inequivalent carbon environments, and therefore the number of distinct NMR resonances, is correspondingly large. Symmetric or repeated subunits, which are structurally reused in the assembly pathway, produce coincident or near-coincident resonances, so the NMR spectrum is not overcounted for repeated fragments. Among carbon types, quaternary carbons (those with no attached hydrogens) carry the most information about molecular topology (these are highly connected and constrained) and accordingly contribute most to the assembly index measurement in NMR. Combining NMR and IR predictions in a weighted average improves the measurement, demonstrating that the two techniques probe complementary features of the assembly space.

These results establish the assembly index as a physical observable in the sense relevant to assembly theory: it is not merely a theoretical construct but corresponds to a measurable feature inferred from the molecule's physical interactions. Measurement does not require full structural elucidation, meaning as metrology improves, the assembly index can be measured even in cases where no mathematical description of a molecule or sample is known. This feature is particularly important in the context of alien life detection, where the structure of molecules for which we have spectral data will likely not be known: in such cases, the experimentally validated theory allows us to directly assess the molecular assembly index from measurement without using structural knowledge as an intermediary.

## Extending Assembly Index Measurement to New Substrates

The experimental validation described above applies to covalent organic molecules. Extending this framework to other substrates requires both a new formalism and careful attention to what can be measured, a distinction that is critical to the generalization of AT to testable hypotheses about measuring of life across diverse substrates. We next describe two substrates where the assembly framework has been expanded, which, in addition to molecules, will inform the general formalism that follows.

### Solid-State Materials

For solid-state materials, including crystals, minerals, and engineered materials, structure can be decomposed hierarchically: elementary bonds define the construction of a unit cell, and the periodic arrangement of unit cells defines the macroscopic object[35]. This yields a two-component assembly index, $a_i = a_\mathrm{u} + a_p$, where $a_\mathrm{u}$ is the assembly index of the unit cell computed from bonds as building blocks, and $a_p$ captures the assembly index of the assembled periodic arrangement of the material sample. Analysis of the American Mineralogist Crystal Structure Database across approximately 4,300 mineral structures shows a broad distribution of unit-cell assembly indices, $a_\mathrm{u}$ , ranging from simple periodic structures near a logarithmic lower bound to heterogeneous, asymmetric structures near a linear upper bound, as a function of the number of bonds in the unit cell[35]. Hazen et al. noted values of $a_\mathrm{u} > 15$ for naturally forming mineral unit cells, and suggested that this invalidates the use of assembly theory for molecular biosignatures[37]. However, as Walker et al.[38] emphasize, minerals lie squarely outside the distribution of the life detection model[5] that used tandem mass spectrometry on organic chemistry to put forth the $a \approx 15$ threshold. The threshold concept cannot be naively transferred from molecular to mineral assembly spaces: one must start with a substrate-specific experimental protocol applicable to minerals. The biosignature threshold of $a \approx 15$ established for covalent organic molecules rests on experimental measurements: the assembly index was confirmed in spectroscopic fragmentation data of molecules that exist in high copy number. For mineral unit cells, crystallographic analysis averages over an entire crystal, and the precise assembly index and copy number of individual unit cells are not directly accessible experimentally in the same way that tandem mass spectrometry accesses individual molecular ions[38]. The hypothesis of assembly theory is that a threshold will exist for *any* substrate and can be interrogated experimentally with an appropriate metrological framework in place, but the value of that threshold is expected to depend on the assembly space, defined by the elementary units, the joining operations, and the physical constraints of the substrate. It is an open question whether or not the threshold will be universal across all substrates[14,38]. For example, in solid state materials, assembly theory formalizes a material technosignature threshold[35] but more work is needed to empirically validate this threshold in mineral assembly.

## Planetary atmospheres

A third example of a substrate-specific application is planetary atmospheric chemistry, motivated by the prospect of remote life detection on exoplanets[36]. Here, rather than probing individual molecules in a laboratory sample, one can ask whether the ensemble of molecules comprising a planetary atmosphere carries a measurable signature of selection at a global scale. The hypothesis is that a biosphere capable of generating and maintaining complex chemistry at a planetary scale will leave a trace in the atmosphere's molecular diversity: the atmosphere will contain molecules whose co-construction requires more recursive steps than the abiotic baseline and will exhibit a deeper reuse of shared intermediate fragments across detected species. Under this hypothesis, an inhabited world's atmosphere should display a higher assembly index and a higher density of shared chemical substructure than a comparable abiotic atmosphere, reflecting the directed exploration of chemical space that selection enables. The measurement scheme extends the application of IR spectroscopy for isolated molecules[7] to mixtures of molecules that appear in atmospheric spectra[36]. Here assembly index is determined from the idealized simulated IR of gaseous mixtures and from the reduced spectra obtained from atmospheric retrieval pipelines, without necessarily requiring knowledge of an atmosphere's full chemical inventory. As with solid-state materials, an atmospheric assembly threshold has yet to be observationally confirmed.

## The Substrate-Dependence of Assembly Thresholds

Taken together, the theory development across molecular, solid-state, and atmospheric substrates suggest paths toward a generalized framework. In each case it becomes clearer how assembly index must be implemented as a substrate-specific quantity, defined relative to an assembly space whose elementary units and joining operations reflect the physical constraints of the substrate under study. The central principles of theory remain the same across these substrates. The assembly threshold, defined as a boundary in assembly space above which objects cannot arise and persist in detectable abundance through undirected processes, is expected to be substrate-specific. For covalent organic molecules, this threshold is empirically established at $a \approx 15$. For crystalline materials and planetary atmospheres, the threshold is a well-defined theoretical target but has not yet been experimentally validated. For other substrates the theoretical frameworks and empirical tests have yet to be developed, in part motivating us to provide a unified view of the current formalism underlying AT in this current work.

Our contention is that any broadly successful theory of physical complexity will necessarily use substrate-specific formalisms. No one formalism could remain physically meaningful and tractable across diverse substrates[39]. AT accomplishes this with general physical principles, but substrate-specific implementation of those principles. A complexity measure that is substrate-independent, such as Kolmogorov complexity[40], achieves universality at the cost of being less interpretable, intractable, and empirically inaccessible. Further, Kolmogorov complexities of finite objects are entirely subject to the choice of reference machine[41]. The molecular assembly index is tractable and does not conform to the computer science concept of universality. In defining a substrate-specific assembly space, the assembly index can be measured from a physical sample, with a metrological referent, rather than only existing solely as a calculated quantity. Thus, even in cases lacking a complete description of an object, one can nonetheless measure its assembly index.

Having established how metrology informs the design of substrate-specific assembly spaces, we can now introduce a formal framework for assembly spaces and assembly indices, which generalizes across substrates. The general framework becomes meaningful when the causal constraints of a physical substrate are encoded into it.

# A Generalized Substrate-Independent Formalization of Assembly Index and Assembly Spaces

Having introduced some of the physical concepts underlying assembly theory, we now turn to our primary goal this work: to provide generalized formal definitions. An ***Assembly Space*** is denoted $\mathbb{A} = (\Omega, J)$, may be defined by a set of *Objects*, $\Omega$, and a ternary relationship over that set, $J: \Omega^3 \rightarrow \{0,1\}$. Let $x, y, z \in \Omega$, then we say that there exists a *Joining Operation* that turns $x$ and $y$ into $z$ if and only if $J(x, y, z) = 1$. Thus, $J$ encodes the set of possible joining operations. We require a unique subset of objects $T \subset \Omega$ called *Units*, which are the product of no joining operation, and generate the rest of $\Omega$ under the finite application of joining operations.

An ***Assembly Path*** can be represented by a finite sequence of object triples, $((x_1, y_1, z_1), \ldots, (x_n, y_n, z_n))$, that satisfy the following property: $\forall\, i \in \{1,2, \ldots, n\}$, $x_i, y_i \in T \cup \{z_1, z_2, \ldots, z_{i-1}\}$ and $J(x_i, y_i, z_i) = 1$. That is, an assembly path is a sequence of joining operations that act on units or products of previous joining operations.

For any set of objects, $X \subset \Omega$, the *Assembly Index* of that set, denoted $a(X)$, is the shortest length of an assembly path that has $X \setminus T$ as its products. If $\{(x_1, y_1, z_1), \ldots, (x_n, y_n, z_n)\}$ is a shortest assembly path for $X$, then $a(X) = n$ and $X \setminus T \subseteq \{z_1, \ldots z_n\}$. When the number of objects is larger than one, e.g., $|X| > 1$, then $a(X)$ is referred to as the *Joint Assembly Index* for the set of objects.

The specifics of objects in $\Omega$, as well as the joining operations that relate them, are determined on a substrate-specific basis, see examples for molecules[5,7,16] and minerals[35]. The formalization of objects and joining operations mirrors relevant constraints of the physical substrate and has physical meaning (see e.g, Walker *et al.*[38] for a discussion of mineral assembly). This substrate-dependence is the most critical distinction between assembly theory and algorithmic complexity: whereas the power of algorithmic complexity comes from its universality and substrate-independence, the power of the assembly index lies in its explicit grounding in the substrate-dependence of physical mechanisms of causation. We will now detail two different substrate-specific assembly space formalisms, as shown in **Figure 1**, for a molecular and polymer example, which illustrate how the general definitions can be embedded in substrate-specific formalisms by encoding metrologically accessible physical constraints in the formal representations.

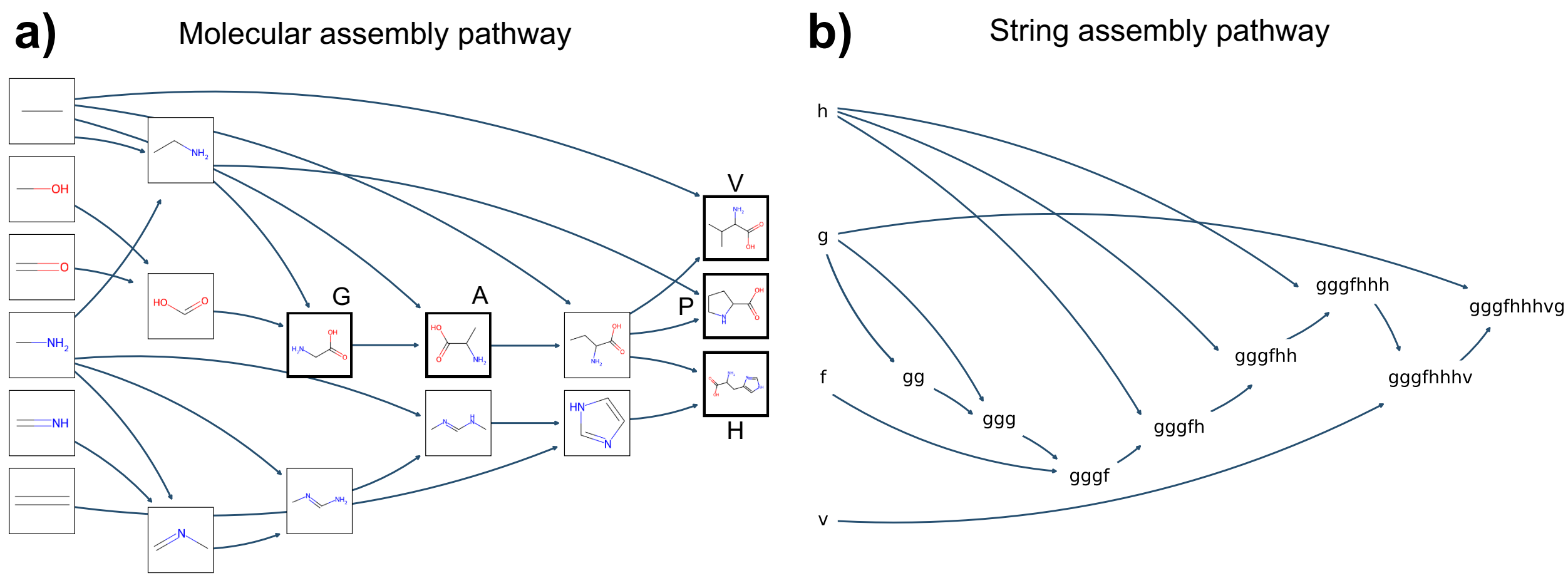

**Figure 1: Assembly spaces for molecules and strings.** A) Joint molecular assembly pathway for the amino acids glycine (G), alanine (A), proline (P), valine (V), and histidine (H). The leftmost nodes represent molecular assembly's elementary units, i.e., molecular bonds. Each joining operation merges two graphs by identifying and contracting specific nodes introducing causal constraints in the form of molecular bonds. B) String assembly path of a short peptide chain composed of G, A, P, V, and H represented as a simple letter chain, joining operations represent peptide bond formation.

## Molecular Assembly Space: $\mathbb{A}_{mol} = (\boldsymbol{\Omega}_{mol}, \boldsymbol{J}_{mol})$

Molecular assembly spaces are described empirically in Jirasek *et al*[7] and Marshall *et al.*[5], with fast algorithms for their computation from molecular graphs presented in Seet *et al*[42]. Here, we represent the objects, $\Omega_{mol}$, in a molecular assembly space with "molecular graphs". A molecular graph is any simple connected graph with at least one edge, where each vertex is labelled with the name of an element ("C", "O", etc.) and each edge is labelled with a type of chemical bond ("single", "double", etc.). The units are the set of molecular graphs with one edge (a single bond). For any three molecular graphs, $x, y, z \in \Omega_{mol}$, we have $J_{mol}(x, y, z) = 1$ if and only if ∃ some sequence of vertices, $\{v_1, \dots, v_n\}$ from $x$ and $\{u_1, \dots, u_n\}$ from $y$, such that $v_i$ and $u_i$ share a label for all $i = 1, \dots n$, and if we contract each $(v_i, u_i)$ pair, we produce a graph isomorphic to $z$. For an example of this contraction process, see **Figure 2.** This formalism is consistent with the molecular assembly spaces presented in recent literature[5,7,22,43].

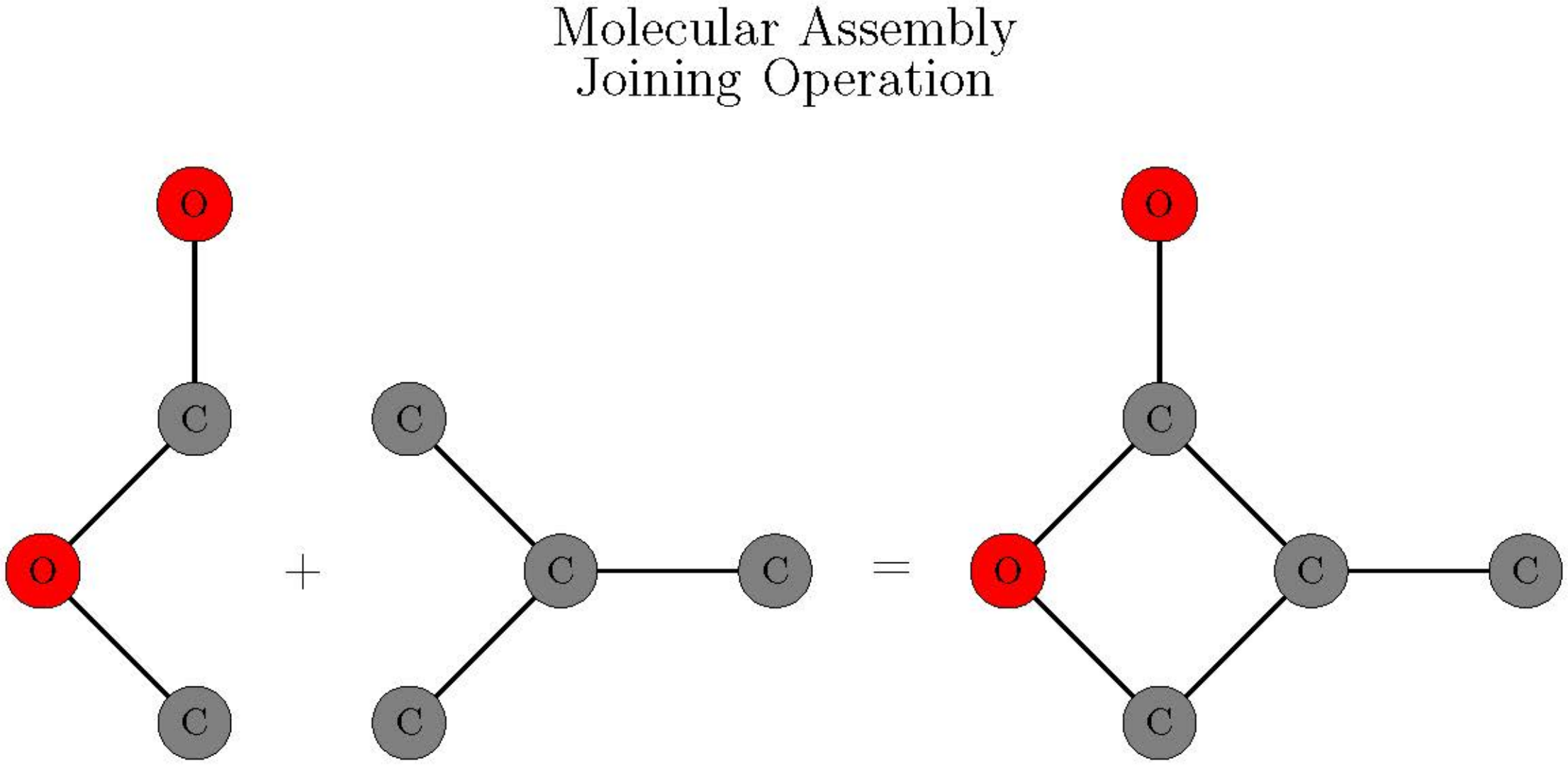

**Figure 2: An example molecular assembly space joining operation.** A joining operation combines two molecular graph fragments to form a new molecular graph. This is done by identifying two carbon atoms from one fragment with two carbon atoms from the other.

## String Assembly Space: $\mathbb{A}_{str} = (\boldsymbol{\Omega}_{str}, \boldsymbol{J}_{str})$

We define a string assembly space with units, $T_{str}$, that are a set of characters, and the objects are the set of strings built from those characters, $\Omega_{\text{str}} = T_{str}^*$. For any three strings, $x, y, z \in \Omega_{\text{str}}$, we have $J_{str}(x, y, z) = 1$ if and only if $z$ can be formed by concatenating a single copy of $x$ and $y$.

This formalism is applicable to systems such as natural language and polymers, where the units could represent characters[44] or monomer species[16,23], see **Figure 1**.

## Representations of Assembly Paths

The formal definition of assembly paths and assembly indices given in the previous section specifies the set of elementary units and a sequence of joining operations starting from them. To connect to the theoretical framework of AT, these must be embedded in a substrate-specific context, with the most extensively developed example being molecular assembly theory[5,7,16]. While the physical concept of a molecular assembly space is well defined by its metrology, across the literature, different representations of the concept have been drawn, notated, and reasoned about, often without explicit acknowledgment that different visual and notational choices preserve different aspects of the underlying physical structure. Some representations show the full ordered sequence of joining operations; others show only the objects produced; still others show only which objects were available at each stage of construction. These are not competing formalisms; they are projections of the same assembly path onto different representations, each emphasizing distinct aspects of that path.

Our goal is to provide an accessible formalism for using assembly paths across contexts; thus, we next present a unified framework for representing assembly paths. Specifically, we identify four representations that appear repeatedly in the literature: the assembly path, the poset path, the object path, and the pool path, see **Figure 3**. These form a lattice, ordered by how much information about the assembly paths is preserved. Alternatively, this lattice describes how these path representations partition the set of assembly paths.

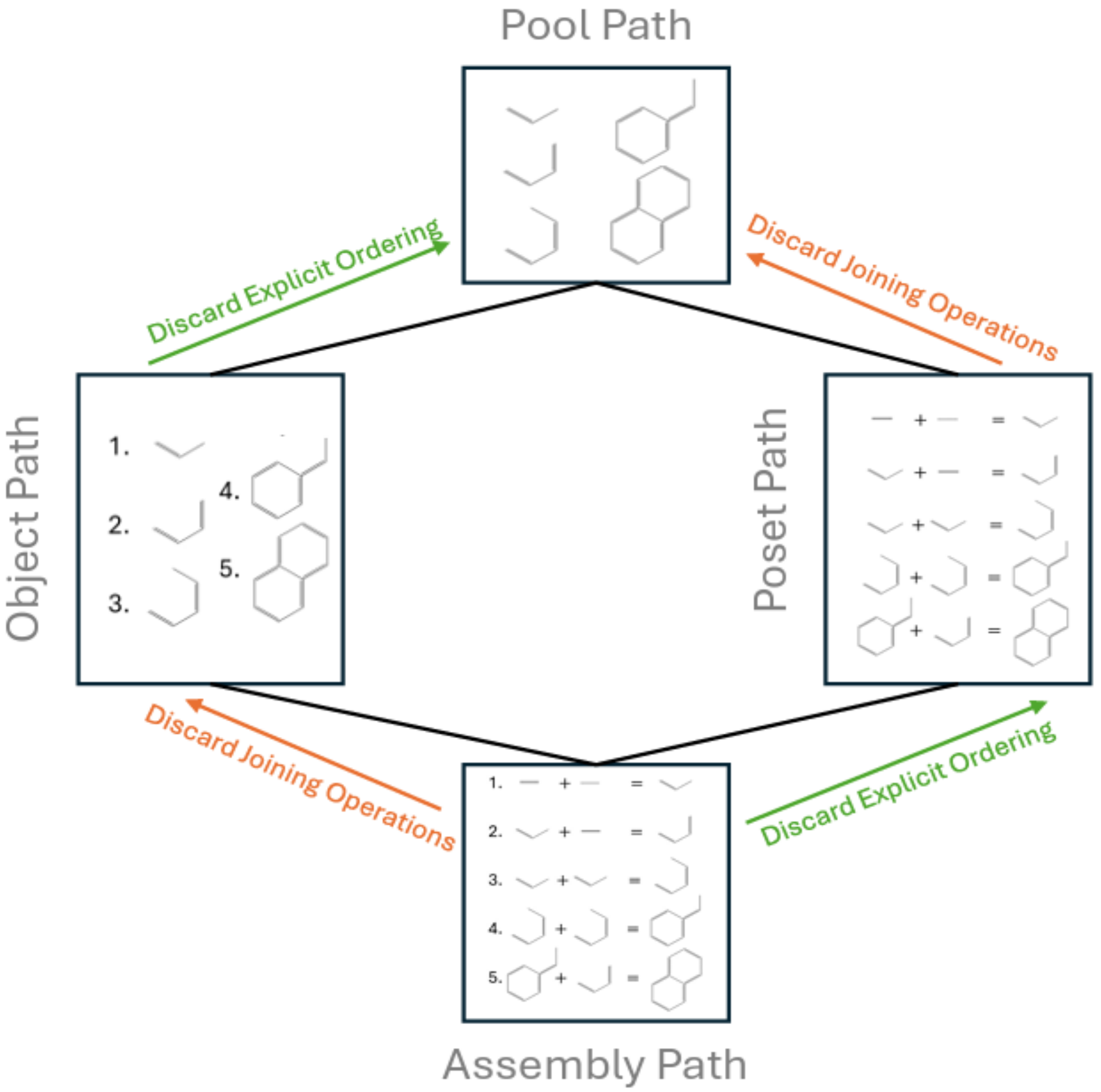

**Figure 3: Lattice depiction of the relationships between four representations of the same assembly path for the molecule naphthalene. Bottom)** An assembly path depiction specifying the full sequence of joining operations. **Center Right)** The poset path includes only the unordered set of joining operations. **Center Left)** The object path includes only the product objects produced by each joining operation along the path. **Top)** The pool path, includes only the unordered set of product objects. Moving up the hierarchy yields representations that are strictly less specific than those at lower levels: object path and poset path representations are at the same level in the hierarchy because their specificities are incomparable.

The four representations in **Figure 3** can be arranged by the amount of structural information they encode about an assembly pathway. What differs across these representations is not the value of the assembly index but what information is made visible about the structure of the underlying assembly space. Different applications of assembly theory may require different representational choices, and clarity about these choices is essential for engaging with, and continuing to develop, the theory's formal framework.

## Assembly paths

The assembly path is the most specific representation we consider, and it was defined in the preceding section. The molecular assembly pathway of *Figure 1a* in Liu et al[16] is notable in representing a full molecular assembly path, and is the only such illustration in the literature to date. Assembly paths are pedagogically useful because they make the step-by-step structure concrete and directly readable.

## Poset paths

A ***Poset Path Representation*** is specified by the finite set of object triples, $\{(x_i, y_i, z_i)\}$, representing the set of joining operations along a valid assembly path. The poset path retains only the causally meaningful content of an assembly path[14]. In the poset path representation, joining operations are ordered only insofar as their input-output dependencies require; any total order consistent with the partial order yields a valid assembly path. This representation is the natural mathematical one for most treatments of assembly spaces and is used in the initial formalism of Marshall et al.[15]. The Hasse-diagram-like depictions of assembly spaces in Sharma et al.[6] are also poset paths: they show the full structure of causal dependencies among joining operations, with the set of possible total orderings implicit in the diagram rather than picking out any one.

The key advantage of the poset representation is that it is the minimal representation that preserves the physical content of the assembly space: it records exactly which operations must precede which others, and no more. Two assembly paths that differ only in the ordering of causally independent steps correspond to the same poset path.

## Object paths

An ***Object Path Representation*** is specified by the finite sequence of objects, $(z_i)_{i=1}^{n}$, produced along an assembly path. The object path retains only the sequence of objects produced during the assembly path, abstracting away the specific joining operations that produced them. This representation appears in literature that emphasizes the intermediate steps in assembly paths, e.g., which objects are created along the way, rather than the detailed mechanics of how each one arises. The object path is appropriate in contexts where the set of intermediate objects is the empirically relevant quantity; for example, when one wants to ask whether a specific substructure or motif appears along the path of a target, without committing to what substructure it was joined to. In chemical applications, the object path corresponds to the sequence of molecular fragments that a path passes through, which is a potential quantity of interest for comparing assembly pathways to known biosynthetic or retrosynthetic routes[16]. The object path retains enough information to reconstruct the assembly index, since the length of the object path equals the length of the underlying assembly path, but not enough to always reconstruct the detailed full assembly space causal structure.

## Pool paths

The ***Pool Path Representation*** is specified by the set of objects, $\{z_i\}_{i=1}^{n}$, produced along an assembly path. The pool path is the least specific among the representations depicted in **Figure 3**. It records only the set of objects produced along the assembly path. The pool path records which objects are produced along the path, but not the specific joining operations that produce them. Pool-path-like representations are uncommon but are included herein for completeness. The pool path representation is the minimal join between the poset and object path representations in the lattice of assembly path representations.

### Relationships within the Path Hierarchy

The lattice in **Figure 3** can also be more formally understood as a commutative diagram of representations, each arrow of which discards a specific piece of information (either explicit order of independent causal steps, the structure of causal steps or both). The space of distinct structures corresponding to each level generally grows moving down the hierarchy: a given pool path will correspond to many poset paths and many object paths, and each poset or object path will correspond to many assembly paths. This structure clarifies representations in the literature: different works may represent the same assembly path of an object while using different representations, and their depictions may appear to differ even when they agree on what matters physically (e.g. the definition of causal joins within the assembly space and the assembly index for the object). Algorithms that calculate the assembly index often use yet another representation that (like all path representations) can be placed into this lattice.

## Formal Grammars and Approximate Methods for Computing Assembly Indices

A formal correspondence between assembly paths and formal grammars for strings was pointed out by Abrahão *et al.*[17] and discussed in Masierak[18]. Molecular assembly index calculations were developed from a double pushout graph rewriting formalism in Flamm *et al.*[20], where their formalism gives the same assembly index but different a structure to paths. Here we show how the generalized (substrate-independent) definition of assembly spaces allows defining the formal correspondence with formal grammars.

Formal grammars are generative symbolic systems that play a central role in theoretical computer science[45]. Formal grammars have been applied beyond text, most notably to graphs[46]. They have widespread utility with applications in data compression[47], structural complexity quantification[48], hierarchical organization elucidation[49], the study of biological polymers[50,51], and building rule-based models of chemistry[52–55,48]. We focus on formal grammars that can generate either strings or graphs. A formal grammar is canonically represented by a tuple, $G = \langle N, T, P, S \rangle$, where $N$ is a set of *non-terminal symbols*, T is a set of *terminal symbols*, P is a set of *production rules*, and S is the *starting symbol* (see **Figure 5** for an example). We intentionally used $T$ to represent the units of an assembly space for illustrating formal correspondence. Though one could construct grammars to represent any level of path hierarchy, there is a canonical definition, and it matches up with the poset path representation. We focus on the poset path representation to describe string assembly spaces in what follows[56].

Given an assembly space, $\mathbb{A}^\dagger = (\Omega^\dagger, J^\dagger)$ , a corresponding class of formal grammars, $\mathcal{G}^\dagger$, can be constructed by identifying the units with terminal symbols, all other objects with non-terminal symbols, and the joining operations with production rules. Each assembly path in $\mathbb{A}^\dagger$ has a corresponding grammar in $\mathcal{G}^\dagger$, whose non-terminals correspond to its products, and which has the same length. Thus, the problem of finding a shortest assembly path in $\mathbb{A}^\dagger$ can be translated to the problem of finding a smallest grammar in $\mathcal{G}^\dagger$, see **Figure 5**.

In general, $N \cap T = \emptyset$ and $S \in N$. Let $(N \cup T)^*$ denote the set of possible symbolic structures built from $N \cup T$. Then, the production rules are maps from $(N \cup T)^*$ into itself. The terminal symbols are preserved by the production rules, but they may be relevant "context", informing the application of the production rules. For example, one could use a production rule to represent a molecular operation that adds an -OH group onto a carbon in a 6-ring. In this case, the 6-ring serves as the context for the operation, since it must be present even though it remains unchanged. The *size* of a formal grammar refers to the number of production rules it possesses. We will define two specific classes of formal grammars, the first applies to strings, and the second to graphs, to demonstrate how the correspondences can be used in substrate-specific contexts.

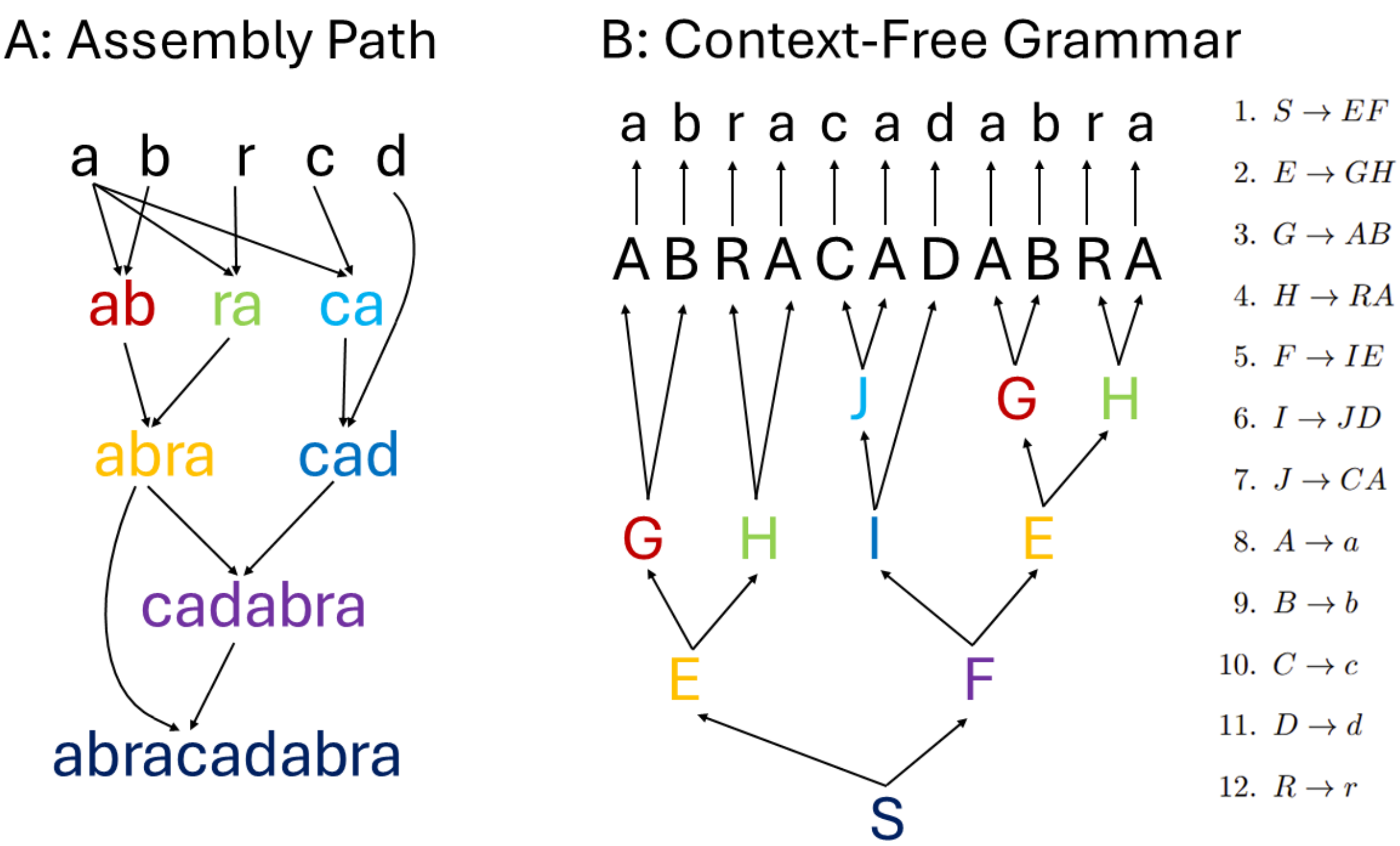

**Figure 4: An assembly path and its representation by a formal grammar. A)** Diagrammatic representation of a minimum assembly path to “abracadabra”. In assembly theory the size and structure of the assembly paths are relevant features of causal structure, not the specific representational labels of the entities along the path. **B)** Diagram of a context-free grammar (CFG) labeling of the assembly path, and a corresponding minimum CFG rule set allowing one to map an assembly path for “abracadabra” onto new labels in a grammar. This grammar builds the string “abracadabra”: it is a *minimum* CFG because no CFG constrained to the same rules corresponding to assembly theoretic causal joins can construct “abracadabra” with fewer rules. The assembly index is $a = 7$.

## Assembly paths in $\mathbb{A}_{str}$ and Context-Free Grammars

Context Free Grammars (CFGs) are a well-studied class of formal grammars. Though one can define context free grammars for graphs and other objects, we will restrict ourselves to strings when discussing CFGs in this paper. Let us represent a CFG by $G_{CFG} = \langle N, T, P, S \rangle$. We will write the non-terminal symbols as capital letters, $N = \{A, B, C, \dots\}$, with the starting symbol being among them, $S \in N$. The terminal symbols will be the alphabet of the strings that we can build, so $T = \{a, b, c, \dots\}$. To match string assembly space in the next section, we will restrict ourselves to CFGs in Chomsky Normal Form[57], meaning that each of our production rules will have one of two forms: $A \rightarrow BC$ or $A \rightarrow a$, for some $A, B, C \in N$ and $a \in T$. In other words, each production rule replaces one non-terminal symbol with either two non-terminal symbols or one terminal symbol. "Context free" refers to the fact that these production rules do not depend on anything other than the non-terminal symbols they replace. We can now map an assembly path in $\mathbb{A}_{str}$ to a corresponding CFG. Let $\{(x_1, y_1, z_1), (x_2, y_2, z_2), \dots, (x_n, y_n, z_n)\}$ be an assembly path of length $n$ in $\mathbb{A}_{str}$. First, let the terminal symbols be the units, so $T = T_{str}$. Now we can define the non-terminal symbols and production rules as we read through the path, from $(x_1, y_1, z_1)$ to $(x_n, y_n, z_n)$. For each $(x_i, y_i, z_i)$ encountered, we generate a new non-terminal symbol for each of $x_i, y_i, z_i$ that has not yet been encountered. So, for $(x_1, y_1, z_1)$ we will generate non-terminals $X_1, Y_1, Z_1 \in N$, but if we encounter $Z_1$ later as $x_i$, we will use $Z_1$ instead of generating $X_i$. Now if $x_i$ or $y_i$ is a unit that has not yet been encountered, then we generate a production rule like $X_i \rightarrow u$ or $Y_i \rightarrow v$, where $u, v \in T$ are characters of the string (are terminal symbols). Finally, we generate a production rule of the form $Z_i \rightarrow X_i Y_i$, though as mentioned earlier, $X_i$ and $Y_i$ may have different labels if we have encountered them earlier. Then we can move on to the next joining operation, $(x_{i+1}, y_{i+1}, z_{i+1})$, and begin again. For simplicity, assume we started with a shortest assembly path to $z_n$, then $Z_n$ is the starting symbol, and our CFG, $G_{CFG} = \langle N, T, P, Z_n \rangle$, is complete. If the assembly path is of minimum length, then the correspondent G is also a smallest grammar. The assembly path length is equal to the number of non-terminal producing rules in $G_{CFG}$.

## Assembly paths in $\mathbb{A}_{\boldsymbol{mol}}$ and Hyper-Edge Replacement Grammars

Hyperedge Replacement Grammars (HRGs) are graph rewrite systems that have seen extensive study[46], with applications to pattern recognition[58], natural language processing[59], and drug discovery[60]. We will define our class of HRGs closely following Chiang *et al.*[59]. First, we must define a *labeled hypergraph* as a tuple $H = \langle V, E, l \rangle$, where $V$ is a finite set of nodes, $E$ is a finite, nonempty sequence of vertices from $V,$ and $l: V \cup E \to C$ assigns a label (drawn from a finite set $C$) to each vertex and each edge. In representations of assembly spaces, one should consider the vertex labels as element names ('Carbon', 'Oxygen', etc.) and terminal edges as bond types ('Single', 'Double', etc.). The labels for non-terminal edges will be distinct but arbitrary. We define our terminal edges to always have degree two and be undirected, so they match molecular assembly space. Next, we define a *hypergraph fragment* as a tuple $\langle V, E, l, X \rangle$, where $\langle V, E, l \rangle$ is a hypergraph and $X \subseteq V$ is a sequence of distinct nodes called the $external\ nodes$.

We now define a HRG as a tuple $G_{HRG} = \langle N, T, P, S \rangle$ where $N$ and $T$ are finite, disjoint sets of non-terminal and terminal edges respectively, $S \in N$ is the start symbol, and $P$ is a finite set of production rules of the form $A \to R$, where $A \in N$ and $R$ is a graph fragment constructed from $N \cup T$. Finally, we describe the HRG rewriting mechanism. Given an HRG $G$, we define the relation $H \Rightarrow_G H'$ as follows. Let $e = (v_1, \dots, v_k)$ be a hyperedge in $H$ with label $A$. Let $(A \to R)$ be a production rule of $G$, where $R$ has external nodes $X_R = (u_i, \dots, u_k)$. Then we write $H \Rightarrow_G H'$ if $l(v_i) = l(u_i)$ for $i = 1, \dots, k$, and $H'$ is the graph formed by removing $e$ from $H$, taking an isomorphic copy of $R$, and identifying $v_i$ with (the copy of) $u_i$ for $i = 1, \dots, k$.

Following the CFG-string case, we will demonstrate how the correspondence can be applied to molecular assembly space and hyper-edge replacement grammars (HRG) by describing a procedure for translating molecular assembly paths into an HRG. Let $\{(X_1, Y_1, Z_1), (X_2, Y_2, Z_2), \dots, (X_n, Y_n, Z_n)\}$ be an arbitrary assembly path in $\mathbb{A}_{mol}$. We will work our way up the assembly path, defining a production rule corresponding to each joining operation; for example, see **Figure 5**. For $Z_1$, define a hypergraph, $Z_1{}'$ formed by joining $X_1$ with a hyperedge, $e_1$, labeled $l(e_1) = 1$, so that the external vertex identified between $e_1$ and $X_1$ is the same vertex identified between $Y_1$ and $X_1$ in the molecular assembly joining operation $(X_1, Y_1, Z_1)$. Finally, we generate a production rule $(e_1 \rightarrow Y_1)$. Now, we can sequentially translate the remaining assembly path. For each $Z_i$, we define a hypergraph, $Z_i{}'$ formed by joining $X_i$ with a hyperedge, $e_i$, labeled $l(e_i) = i$ so that the sequence of external vertices where $e_i$ and $X_i$ overlap is the same as the sequence of vertices identified in the molecular assembly joining operation. If $X_i = Z_j$ for some $j < i$, we replace $X_i$ with the hypergraph $Z_j{}'$. Likewise, we generate a production rule of $(e_i \rightarrow Y_i')$ where $Y_i{}' = Z_j{}'$ if $Y_i = Z_j$ for some $j < i$, else $Y_i{}' = Y_i$. So now our translated HRG reads $G = \langle N, T, P, S \rangle$ where $N = \{e_1, \dots, e_n\}$, $T$ is the set of terminal symbols, $P = \{(e_i \rightarrow Y_i'): i = 1, \dots, n\}$, and $S = Z_n$. The number of production rules is equal to the number of joining operations; thus, the fewest number of distinct production rules required by this kind of HRG to produce a given molecular graph is equal to that graph's assembly index.

Importantly, these HRGs do not have a full structural equivalence with molecular assembly paths, see **Figure 5**. For one, the order in which production rules are applied can sometimes be permuted. Also, the ordering of external vertices can sometimes be permuted in cases where some fragments have nontrivial automorphisms, see Rules A and B in **Figure 5** for an example. Further, it was arbitrary to make the $Y_i{}'$s into hyperedges; for example, consider Rules A and C in **Figure 5**. This demonstrates how, in general, it is non-trivial to go from the generalities of formal grammars to substrate-specific assembly spaces. However, given the general correspondence, one could define another class of HRGs that mod out by these equivalences, which would possess a one-to-one correspondence with molecular assembly paths, being more complicated in its introduction of more causal information from the structure of the assembly space.

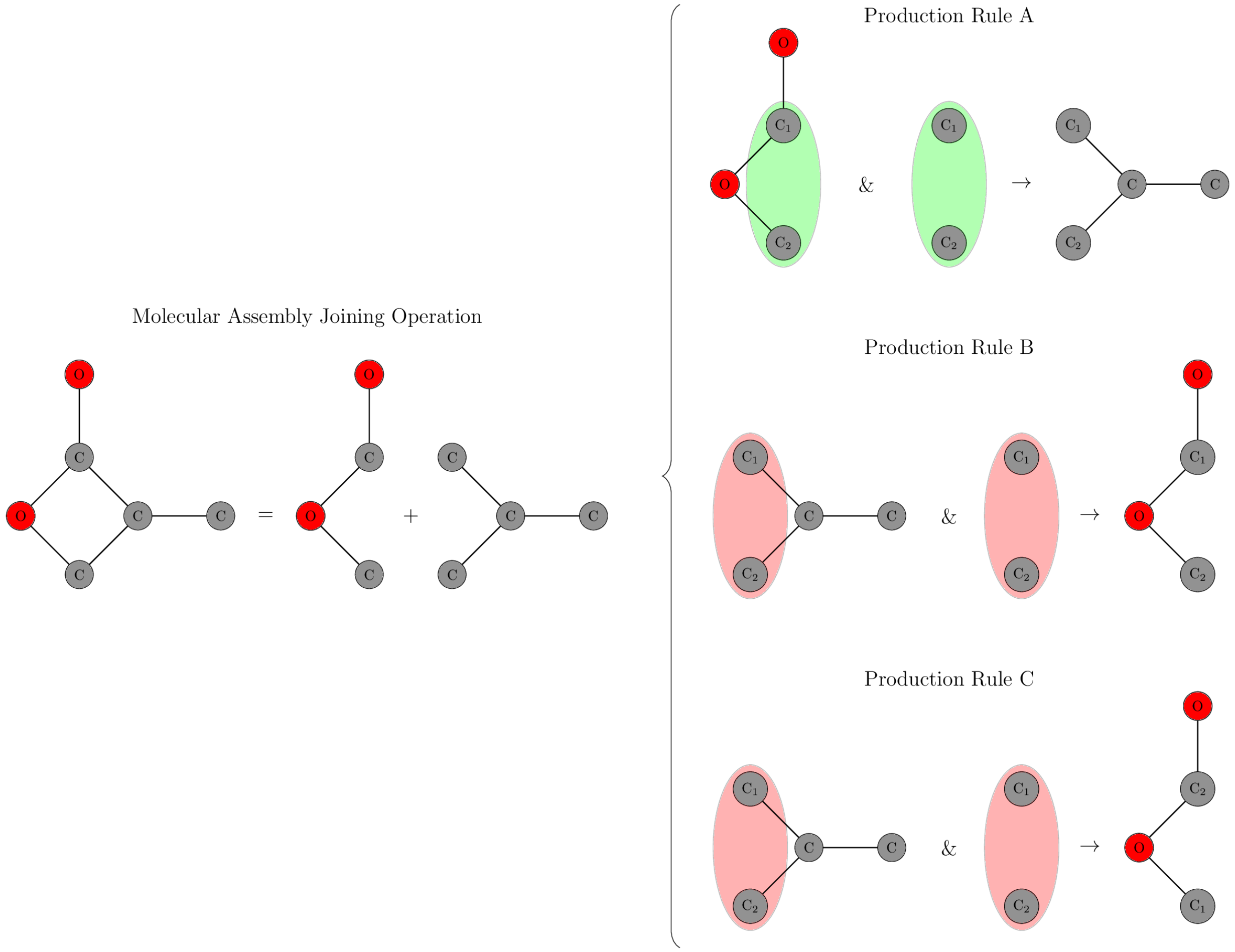

**Figure 5: The molecular assembly space joining operation on the left corresponds to several canonically distinct Hyper-edge replacement grammar (HRG) production rules, demonstrating how it is often non-trivial to represent assembly spaces as grammars.** Graphs represent molecular fragments. Each HRG production rule shown here corresponds to this joining operation when acting upon the hypergraph to its left. Rule A is distinct from B and C as it replaces its hyperedge with a distinct graph. Rules B and C are distinct from one another because they differ in the ordering of their external vertices. This multiplicity of corresponding production rules is an artifact of the canonical definition of an HRG, and it stops us just short of full structural equivalence to the causal structure encoded in the assembly space. However, this class of HRGs

does at least have the property that the size of a smallest grammar is equal to the molecular assembly index, allowing expanded algorithmic approaches to assembly index computation from molecular graphs.

## Fast Upper and Lower Bounds on Assembly Indices

The correspondence can be used to repurpose existing algorithms to efficiently computate the assembly index. Computing the assembly index is generally difficult[19], so methods to efficiently bound or estimate assembly index can have great utility. We apply several pre-existing grammar algorithms to the problem of estimating the assembly index by leveraging the correspondence between assembly paths and formal grammars. We compare the performance of these algorithms to the state-of-the-art exact assembly index calculator, AssemblyCPP, which is described in detail in Seet *et al*.[42]. These algorithms have been implemented in the public Python package AssemblyTheoryTools, a central repository for many diverse workflows pertaining to assembly theory (see https://pypi.org/project/assemblytheorytools/). We note that the algorithms presented are not directly applicable to life detection as they are applied to strings; in extending any grammar-based approximations to molecular graphs, lower bounds are likely to lead to poor and/or inaccurate classification and the upper bound is expected to be of more direct applicability.

### Upper Bound

RePair is an algorithm developed to efficiently find approximately minimum CFGs[61]. It operates by turning the most common repeated pair of adjacent symbols in a string into a new non-terminal symbol repeatedly until no repeated symbol pair remains, completing in a time that is polynomial in string length, as shown in **Figure 6**. Given a specific input string, RePair finds a concise CFG encoding. When implemented using the correspondence, it also finds a short assembly path and, therefore an upper bound to the assembly index for strings. **Figure 6** illustrates the upper bound to assembly index; this indicates that the algorithm provides a fast and reliable method for estimating string assembly indices.

Though RePair is generalized to graphs[62], we can also apply the string-native version to molecular graphs as follows: we partition the molecular graph into trails (A trail is a connected graph where all but two vertices have two adjacent neighbors, and those two have just one. They can be drawn as a straight line.), and we apply RePair to the resultant set of 'molecular strings'. The validity of the upper bounds produced this way follows from the observations that (1) the size of any valid assembly path places an upper bound on the assembly index, and (2) first assembling a set of disjoint trails and then combining them into a final molecule is a valid assembly path (though often not minimal). This method yields quick upper bounds for the molecular assembly index. This algorithm effectively restricts us to assembly paths that build these molecular strings before combining them into the full molecule. This is a significant constraint in well-connected molecular graphs, and the resulting assembly paths are not close approximations to the true molecular assembly index. However, for 'stringy' molecules like lipids, this algorithm offers a decent approximation to the true assembly index, as shown in **Figure 7**, enabling its approximation for large polymeric molecules. Future work may aim to extend these results by applying native graph grammar methods.

## Lower Bound

Prior work has shown how lower bounds on the size of a minimum CFG can be attained from a Lempel-Ziv (or LZ) algorithm[63]. The correspondence allows this same algorithm to produce lower bounds to the assembly index of strings in a time that scales as string length squared, see **Figure 6**. The gap between this lower bound and the assembly index is shown for a generic population of strings in **Figure 6**; the lower bound typically tracks the assembly index, offering a fast and accurate means to approximate the assembly index of a string. We note LZ compression is not the same as assembly index as claimed by some authors, but here we implemente only as a lower bound for strings, see e.g. Kempes *et al.* for more discussion[19]

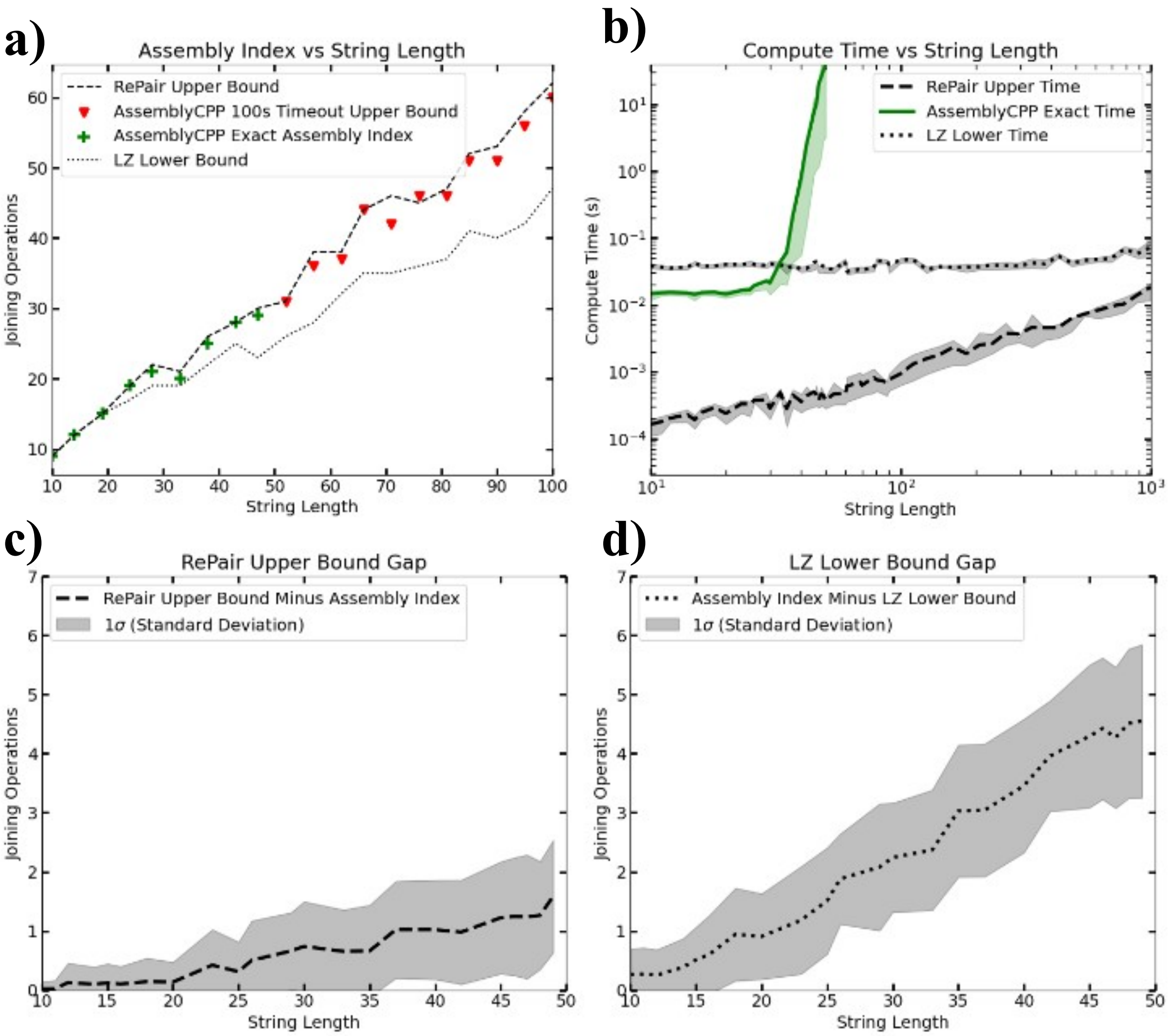

**Figure 6: Fast estimation of upper and lower bounds on assembly index for strings. a)** Random sample of individual 4-character strings with unique lengths between 10 and 100 characters. For each string, the exact assembly index calculated by AssemblyCPP is shown in green, and the upper bound output if it did not finish in 100s is shown in red. The RePair-derived upper bounds are the dashed line, and the LZ-derived lower bounds are the dotted line. **b)** Compute times of these algorithms on a larger ensemble of random 4-character strings. AssemblyCPP is shown in green, while the upper- and lower-bound algorithms are represented by dashed and dotted

lines, respectively; the boundaries of the shaded regions correspond to the first and third quartiles. The AssemblyCPP line is truncated at a string length of 54, where 50 percent of calculations did not terminate within the allotted 100s; the scaling of AssemblyCPP appears less steep than reality due to post-selecting strings whose assembly index calculation finished within 100s. Both bounding algorithms display superior time scaling to the exact calculation, with our implementation of the LZ-derived lower bound having the most favorable scaling.**c)** The gap between RePair-derived upper bounds and exact assembly index for a sample of random 4-character strings; the dashed line is the mean, and the shaded region is one standard deviation around it. **d)** The gap between the exact assembly index and the LZ-derived lower bound for a sample of random 4-character strings; the dotted line represents the mean, and the shaded region indicates one standard deviation around it. Both algorithms reliably track the assembly index, with RePair following it more closely. Given a specific population of strings, these upper and lower bounds can be used to produce fast, accurate estimates, of the assembly index.

## Vector Addition Chain Lower Bounds

Vector addition chains can be considered a formal grammar over vectors of integers of a given dimension. The terminal symbols are the Cartesian basis vectors ([1, 0, 0, ...], [0, 1, 0, ...], ...), and the production rules are vector addition. If the dimension of the vectors is one, then we refer to this grammar as an integer addition chain. As noted previously[15], both vector addition chains[64] and integer addition chains[65] can place lower bounds on the assembly index of a potentially wide range of assembly spaces. These bounds apply to string assembly space and molecular assembly space. The connection to addition chains has also been noted via CFGs[63]. We show the integer addition chain lower bound applied to molecules in **Figure 7**.

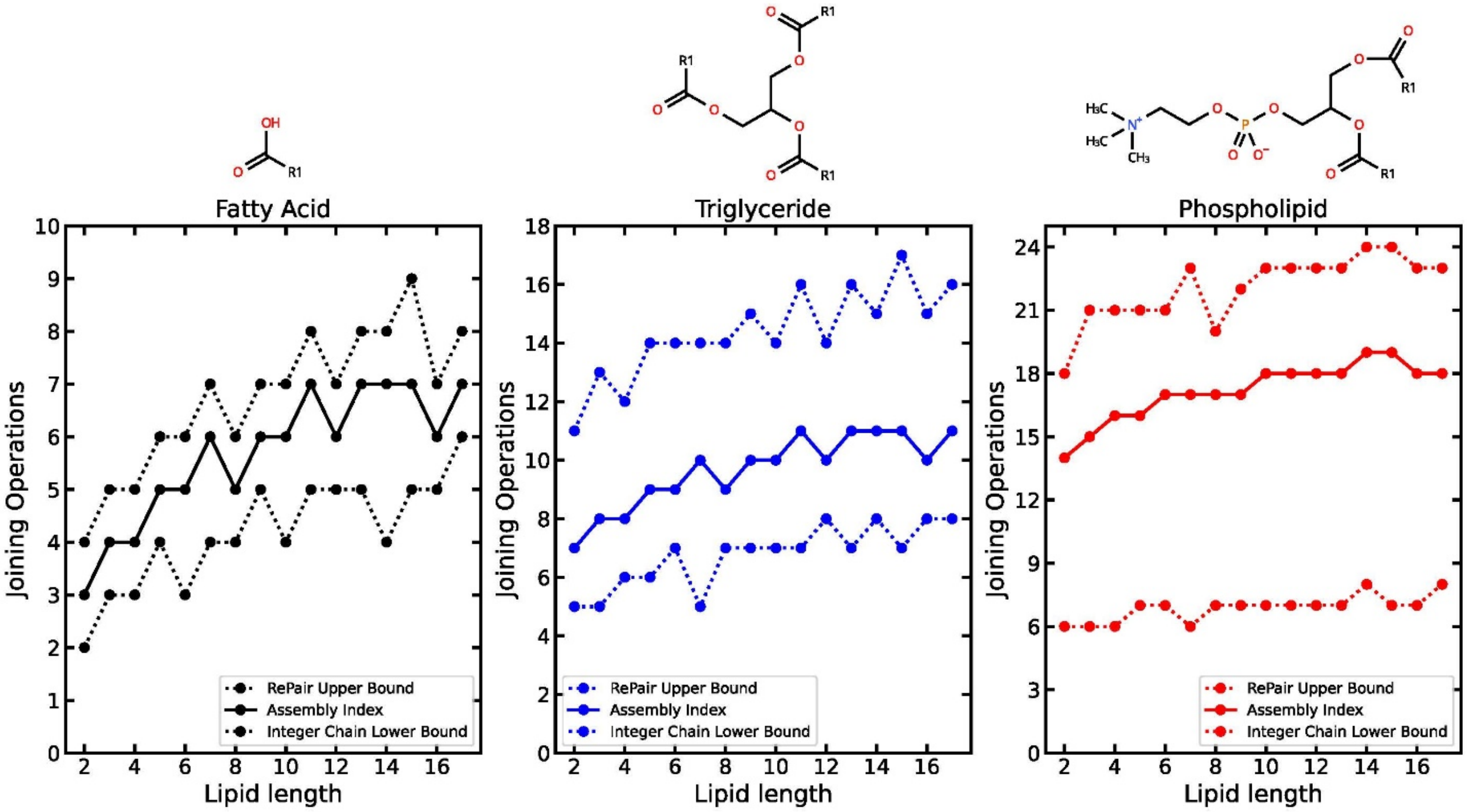

**Figure 7: Bounding assembly index calculations for 'string-like' molecules.** In black, the left panel represents an unbranched fatty acid, in blue, the central panel represents a triglyceride, and in red, the right panel represents a phospholipid with a Phosphatidylcholine head group. The lipid length denotes the number of carbon atoms in the lipid tails, denoted by $R1$ in the molecular diagrams. The exact assembly index is shown as a solid line in each panel, while the addition chain lower bounds and the RePair derived upper bounds are dashed lines. The upper bound is tightest when the molecular graph is very sparse. The upper bound degrades when the head group is more complicated. While string approximations are in general not good approximations of molecular assembly index, for 'string like' molecules they provide reasonable bounds.

Computing the assembly index of arbitrarily large objects is rarely possible (**Figure 6)**, and string approximations to non-polymeric molecules will be poor (**Figure 7**). The computation time for the exact calculation of assembly index (using AssemblyCPP) scales exponentially with string length, while the LZ lower bound is polynomial. Although computing exact values for the assembly index is NP hard, the Repair upper bound and LZ lower bound allow efficient bounding for strings. Thus, using these algorithms to bound the value of the assembly index for strings may be useful in some contexts where AT might be applied.

# Discussion

Each substrate-specific assembly space can be shown to correspond to some class of formal grammars. Through this correspondence, pre-existing grammar compression algorithms can be repurposed to bound assembly indices in the corresponding assembly space, as we have demonstrated here. For strings, the LZ-derived lower bound and the RePair-derived upper bound can together provide a fast and accurate means of estimating the assembly index. Given a particular population of strings, one can utilize distributions analogous to those shown in **Figure 6** to produce robust bounds of assembly index in regimes where exact calculation is inaccessible.

The string bounds we present here are not directly relevant to the problem of molecular life detection[19]. Our hope is that by providing fast methods for bounding assembly index, and clarifying the ontology behind the formalism, researchers will be able to expand the application of the framework. We also note that the correspondences presented herein can also offer further clarity on the relationship between assembly theory and algorithmic complexity as also explicated previously by Kempes et al.[19]. Assembly theory and algorithmic complexity approaches were developed for entirely different purposes[14,40], with only a weak similarity in how both implement a concept of minimality as foundational to their respective concepts of complexity. For finite data, as characterizes any physical system, any assessment of algorithmic complexity will depend entirely on the choice of reference machine, a choice which is made somewhat arbitrarily, which we explore more in future work. By contrast, the formalisms in assembly theory are developed to incorporate knowledge of physics in such a way that we can, a priori, expect reasonable statements about physical complexity, with empirically testable hypotheses that can be explored in the lab.

**Acknowledgments**

We thank Ian Seet, Keith Patarroyo, Juan Carlos Morales Parra, and Leroy Cronin for their comments on an early draft of the manuscript.

**Ethical Statement**

NA

**Funding Statement**

This work was supported by Schmidt Science grant number G-2365174 awarded to SIW and NSF award number 2233001 awarded to GS.

**Data Accessibility**

See https://github.com/ELIFE-ASU/CFG.git for version history of code used to generate data and figures, https://pypi.org/project/assemblycfg/ for an easily installable current version, and https://doi.org/10.5281/zenodo.20562899 for an archived release.

**Competing Interests**

We have no competing interests.

**Authors' Contributions**

G.S. conceptualization, formal analysis, software, methodology, validation, writing – original draft, funding acquisition; R.C. conceptualization, formal analysis, software, methodology, writing – original draft; L.S. software, methodology, validation, writing – original draft, project management; S.W. conceptualization, investigation, methodology, validation, project supervision, resources, management, writing – original draft, review and editing, funding acquisition.